\documentclass[a4paper,fleqn,usenatbib]{mnras}

\usepackage[T1]{fontenc}
\usepackage{ae,aecompl}

\usepackage{xcolor,colortbl}
\usepackage{caption}
\usepackage{soul}

\usepackage{graphicx}	
\usepackage{amsmath}	
\usepackage{amssymb}	

\definecolor{Gray1}{gray}{0.9}
\definecolor{Gray3}{gray}{0.7}
\definecolor{Gray5}{gray}{0.55}
\definecolor{Gray7}{gray}{0.4}
\definecolor{Gray9}{gray}{0.3}
\definecolor{Gray11}{gray}{0.2}
\definecolor{Gray}{gray}{0.85}
\definecolor{White}{gray}{1}

\newcolumntype{a}{>{\columncolor{Gray}}c}
\newcolumntype{b}{>{\columncolor{white}}c}

\title[Topology in Non-Standard Cosmologies]{Large-Scale Structure Topology in Non-Standard Cosmologies: Impact of Dark Sector Physics}

\author[A. L. Watts et al.]{
Andrew L. Watts,$^{1}$\thanks{E-mail: a.watts@physics.usyd.edu.au}
Pascal J. Elahi,$^{2,1}$
Geraint F. Lewis$^{1}$
and Chris Power$^{2}$
\\
$^{1}$Sydney Institute for Astronomy, School of Physics, A28, The University of Sydney, NSW, 2006, Australia \\
$^{2}$International Centre for Radio Astronomy Research, University of Western Australia, 35 Stirling Highway, Crawley, WA, 6009, Australia
}

\date{Accepted XXX. Received YYY; in original form ZZZ}

\pubyear{2016}

\begin{document}
\label{firstpage}
\pagerange{\pageref{firstpage}--\pageref{lastpage}}
\maketitle

\begin{abstract}
Even as our measurements of cosmological parameters improve, the physical nature of the dark sector of the universe largely remains a mystery. Many effects of dark sector models are most prominent at very large scales and will rely on future galaxy surveys to elucidate. In this paper we compare the topological properties of the large scale dark matter distribution in a number of cosmological models using hydrodynamical simulations and the cosmological genus statistic. Genus curves are computed from $z=11$ to $z=0$ for $\Lambda$CDM, Quintessence and Warm Dark Matter models, over a scale range of 1 to $20 h^{-1}$Mpc. The curves are analysed in terms of their Hermite spectra to describe the power contained in non-Gaussian deformations to the cosmological density field. We find that the $\Lambda$CDM and $\Lambda$WDM models produce nearly identical genus curves indicating no topological differences in structure formation. The Quintessence model, which differs solely in its expansion history, produces significant differences in the strength and redshift evolution of non-Gaussian modes associated with higher cluster abundances and lower void abundances. These effects are robust to cosmic variance and are characteristically different from those produced by tweaking the parameters of a $\Lambda$CDM model. Given the simplicity and similarity of the models, detecting these discrepancies represents a promising avenue for understanding the effect of non-standard cosmologies on large-scale structure.
\end{abstract}

\begin{keywords}
\textit{(cosmology:)} large-scale structure -- \textit{(cosmology:)} dark energy
\end{keywords}



\section{Introduction}

Beginning with the measurement of high-redshift Type Ia supernavoe \citep{1999ApJ...517..565P, 1998AJ....116.1009R}, several independent lines of observational evidence have indicated that the expansion of the universe is accelerating at the present day [e.g. baryon acoustic oscillations \citep{2011MNRAS.418.1707B} and weak lensing surveys \citep{2007MNRAS.381..702B}].  Calculations of the energy content of the universe \citep{2016A&A...594A..13P} have shown that we live in a (nearly) flat universe where matter makes up only $\sim30\%$ of the energy budget, most of which is in the form of non-baryonic dark matter (DM). The remaining energy, known as dark energy (DE), provides the negative pressure necessary to drive the acceleration of the universe. The $\Lambda$CDM concordance model posits that dark energy can be explained by a cosmological constant $\Lambda$ that represents a uniform vacuum energy density with equation of state $\rho = -p$. Dark matter is thought to be composed of non-relativistic particles which interact with baryonic matter exclusively through gravity. The enormous success of this model in explaining such features of cosmology as large-scale structure \citep{2015ApJS..219...12A}, and CMB anisotropies \citep{1999Sci...284.1481B} has led $\Lambda$CDM to become the Standard Model of Cosmology. 

Despite its many observational successes, the $\Lambda$CDM model suffers from a number of theoretical weaknesses, namely a \textit{fine-tuning} problem and a \textit{coincidence} problem \citep{1989RvMP...61....1W}. Fine-tuning refers to the fact that if the vacuum energy density $\Lambda$ is constant, its initial value has to be highly fine-tuned in order for the accelerated expansion of the universe to happen late enough for galaxies and stars to form first. The coincidence problem concerns why matter and dark energy have densities of the same order today given that for most of the universe's history one or the other should be clearly dominant. This has motivated theorists to posit either modified gravity formulations, or a dynamical form of dark energy whose strength varies with cosmic time and possibly interacts with other forms of energy in the universe \citep{2015PhR...568....1J}. Given that baryonic matter shows little evidence of non-gravitational interaction with the dark sector, it is interesting to consider models involving the exchange of energy between dark matter and dark energy via e.g. DM particle annihilation. These ``Coupled Dark Energy" (CDE) models can naturally address the theoretical problems for certain classes of interaction potentials which produce late-time accelerated expansion for initial values that vary by several orders of magnitude \citep{2004ApJ...605...21C}.

$\Lambda$CDM has also faced problems in reproducing observed structures on sub-galactic scales, for example the number of satellite dwarf galaxies around galaxies such as our own is far below that expected from $\Lambda$CDM numerical simulations~\citep{2001ApJ...556...93B}. This has led to interest in Warm Dark Matter (WDM) models, where the relativistic velocities of the dark matter particle suppress structure formation below a certain scale, thus producing fewer satellites and different halo formation properties but leaving the large-scale structure intact \citep{2011ApJ...739...38P}. WDM has been shown to leave a surprising imprint on the smooth matter distribution at cluster scales which could subtly affect the topology \citep{2016ApJ...826..212M}. Though solutions to $\Lambda$CDM's small-scale problems may ultimately be found in ultra-faint galaxies \citep{2015ApJ...806...95S} or improved simulations \citep{2016ApJ...827L..23W}, it is interesting to consider the cosmological effect that potential WDM particles might have.

Extensive galaxy surveys have shown that on very large scales, the universe has the structure of a `cosmic web' consisting of knots, sheets and filaments of galaxies and their associated dark matter halos. This structure is the result of gravity magnifying perturbations in the density field that were seeded by primordial fluctuations in the very early universe. The analytic and numerical predictions of the $\Lambda$CDM model are largely in agreement with observations of large-scale structure. An interesting and useful tool for classifying large-scale structure is the \textit{topology} or interconnectedness of underdense and overdense regions \citep{1986ApJ...306..341G}. 

Several topological methods in exist in various dimensions including Minkowski functionals, level-crossing statistics, density extrema statistics, the genus statistic -- each described in \citet{2003ApJ...584....1M} -- and network approaches \citep{2016MNRAS.tmp..603H}. In particular the genus statistic has been applied extensively to observational and simulated galaxy surveys \citep{2014ApJ...796...86P,2013ApJS..209...19C,2015ApJ...799..176S,2012ApJ...751...40J}, hydrogen distributions during reionisation \citep{2008ApJ...675....8L, 2015aska.confE..33W} and studying the CMB \citep{2007MNRAS.377.1668G}. An analytic form for the genus statistic in the case of a pure Gaussian Random Field (as predicted by inflationary theories) has been derived in \citet{1986ApJ...304...15B} and so deviations from this form provide information about non-linear gravitational processes and even the primordial fluctuations from which large-scale structure evolved.

In this paper, we use the genus statistic to measure the topology for a suite of DM-only simulations based on various cosmological models. Our aim is to identify discrepancies between the models during the course of cosmic evolution and over a wide range of length-scales. In Section \ref{sec:method}, we discuss our analysis methods and the details of our numerical simulations. In Section \ref{sec:results} we discuss the results of our topological comparison. Sections \ref{sec:discussion} and \ref{sec:conclusion} describe avenues for further research and the consequences of these results for future observational efforts.

\section{Numerical Methods}
\label{sec:method}

\subsection{Topology of Large-Scale Structure}

The large-scale structure of the universe can be studied mathematically in various ways. The most basic and fundamental is the two-point correlation function which gives the power spectrum: the amplitude of density fluctuations as a function of the scale. Although the power spectrum is useful in differentiating models and studying phenomena such as baryon acoustic oscillations, it only gives lower-order information about the density field and throws away a lot of information about the morphological properties of the matter distribution in the universe. In this paper we study these properties by measuring the topology of the universe in terms of the genus statistic.  

When discussing topology, we are referring to the topology of 2D isodensity surfaces that divide a 3D volume of space into regions above and below a certain density threshold. This technique treats underdensities and overdensities equivalently, focussing on the boundary surface between those regions. In qualitative terms, the genus of a surface S can be defined as 
\begin{equation}
\label{genusdef}
g_S = \text{number of holes} - \text{number of isolated surfaces} + 1. 
\end{equation}
A sphere has no holes and 1 isolated surface, and therefore a genus of $0$; a torus has one hole and 1 isolated surface, and therefore a genus of $1$. The genus of large-scale structure varies in a qualitatively predictable way with the density threshold. For instance, at a low threshold this technique will form isolated boundary surfaces around the most underdense voids and hence give a negative genus number (for several unconnected spheres). Similarly, at a high density threshold everything except large clusters will be excised and the boundary surface will again topologically resemble multiple spheres centred on the clusters. At an average density, the overdense and underdense regions of large-scale structure interlock in a `sponge-like' structure of filaments, tunnels, voids and clusters. The interlocking structure results in many holes but few isolated surfaces so this surface has a high genus number.

The computation of the genus statistic first involves mapping the continuous density field onto cubic grid cells and smoothing the distribution with a Gaussian filter.
The density threshold is then calculated by sorting the three-dimensional density values into a one-dimensional monotonically increasing array and choosing a density value that will excise a given fractional volume of space. If we let $\nu$ parameterize the density threshold, the fractional volume is given by

\begin{equation}
\label{fracvol}
v_f (\nu) = \frac{1}{\sqrt{2\pi}} \int_{\nu}^{\infty} \, \mathrm{e}^{-t^2/2} \mathrm{dt} \equiv \frac{1}{2} \, \mathrm{Erf}_c \Big( \frac{\nu}{2} \Big),
\end{equation}
where $\mathrm{Erf}_c$ is the conjugate error function. The value of the density threshold defines which cells are above and below the boundary surface, and can be calculated by converting the 3D density field into a 1D ordered array and moving along the array until the correct $v_f$ has been excised. The threshold is therefore given by the value in the $n^{\mathrm{th}}$ cell where

\begin{equation}
\label{critdens}
n = \text{floor}[(1 - v_f)N],
\end{equation}
and N is the total number of grid cells. The genus of the boundary surface between high-density and low-density regions is calculated according to the Gauss-Bonnet thereom which connects the topology of a polyhedral surface to the sum of the angle-deficit at each of the vertices, $D_{vertex} = 2\pi - \sum_i V_i$ where $V_i$ are the angles around the vertex . In practice, the angle deficit is calculated with reference to a pre-computed look-up table for all possible vertex configurations. The genus of a surface $S$ with Gaussian curvature $K$ is given by 
\begin{equation}
g_S - 1   = -\frac{1}{4\pi}\int_S K dA = -\frac{1}{4\pi} \sum_{\mathrm{vertices}} D_{vertex}.
\end{equation}
For more details of the genus computation, see \citet{1986ApJ...306..341G}. Given that the genus for the surface is the simple sum of the deficits at the vertices, this compuation can be highly parallelized. This technique assumes periodic boundary conditions, as is the case in all of our simulations. For the patches of the sky used in observational galaxy catalogues care must be taken at the boundaries of the survey regions \citep{2007MNRAS.375..128J} and to correct for systematic measurement errors \citep{2012ApJ...751...40J}.

For a Gaussian random field, the analytic formula for the genus curve is well-known \citep{1986ApJ...304...15B}: 

\begin{equation}
\label{genuscurve}
g_{\mathrm{GRF}}(\nu) = A (1-{\nu}^2)\mathrm{e}^{-{\nu}^2/2},
\end{equation}
where $A$ is the amplitude of the genus curve, related to the power spectrum.
This form is conserved during linear growth, meaning that primordial non-Gaussianity can still be detected today \citep{1986ApJ...306..341G}. Deviations from the Gaussian curve can be analysed using a number of mathematical techniques including Hermite functions \citep{2012ApJ...751...40J} and Betti numbers \citep{2013JKAS...46..125P}.
Following \citet{2012ApJ...751...40J} we decompose the curves further in an orthogonal basis of Hermite functions:

\begin{equation}
g(\nu) = \displaystyle\sum_{n=0}^{\infty} a_n \psi_n(\nu) \\
\implies a_n = \int_{-\infty}^{\infty} g(\nu) \psi_n(\nu) d\nu,
\end{equation}
where the Hermite functions $\psi_n(\nu)$ are weighted analogues of the Hermite polynomials: 

\begin{align}
\psi_n(\nu) &= \frac{1}{\sqrt{n!\sqrt{2\pi}}}\exp^{-v^2/4} H_n(\nu), \\
H_n(\nu) &= (-1)^n \exp^{v^2/2} \Big( \frac{d}{dv} \Big)^n \exp^{-v^2/2}.
\end{align}
The coefficients are calculated using a publicly available Markov Chain Monte Carlo (MCMC) parameter estimation algorithm \citep{2013PASP..125..306F}\footnote{emcee: http://dan.iel.fm/emcee}. We also normalize the coefficients to express the size of their contribution to the spectrum, $\tilde{a}_n = a_n/\sum_m |a_m|$. In this formalism, a pure Gaussian random field would have $|\tilde{a}_2|=1$. The strength of the other modes quantifies the non-Gaussian distortions in the density field. In particular, odd-numbered modes produce an asymmetry in the curve that represents a relative overabundance of clusters to voids or vice versa. Higher order modes tend to have more power in the wings of the curve and hence have less affect on structure at average densities. \citet{2003ApJ...584....1M} also calculated approximations for weakly non-linear gravitational evolution, which occurs below a scale of about $10 h^{-1}$Mpc. To first order in the mass variance $\sigma_0$ 

\begin{equation}
\label{eq:wnl}
\begin{split}
g (\nu) \approx -  A \mathrm{e}^{-{\nu}^2/2}\Big\{ &H_2(\nu) + \Big[ \frac{S^{(0)}}{6}H_5(\nu)\\
& + S^{(1)}H_3(\nu) + S^{(2)} H_1(\nu) \Big] \sigma_0 \Big\},
\end{split}
\end{equation}
where $S^{(0,1,2)}$ are the skewness parameters as defined in \citet{2003ApJ...584....1M}. 

\subsection{Non-Standard Cosmological Models}
\label{sec:alt}

Although the $\Lambda$CDM model of cosmology has been very successful, several extensions to it have been developed. In this work we explore an alternative form of dark matter, Warm Dark Matter, and an alternative form of dark energy, Quintessence. We also tweak the parameters of a $\Lambda$CDM model to confirm whether the differences are due to the effects of the underlying cosmological model.

\subsubsection{$\Lambda$ Warm Dark Matter}
\label{sec:wdm}

Warm Dark Matter particles move at relativistic speeds and have large free streaming lengths. This leads to a `smearing' of small-scale structure and suppression of power at sub-galactic scales \citep{2001ApJ...556...93B,2014MNRAS.444.2333E}. In analysing this model we hope to show that there is no topological difference at larger scales and also to provide estimates on the noise inherent in the genus calculation. A number of WDM candidates exist including thermal relics from the early universe or products of non-equilibrium decay. Observations of the Lyman-$\alpha$ forest at high redshift have constrained the lower limit of the energy of a thermal relic WDM particle to above 3.3 keV \citep{2013PhRvD..88d3502V} at the $2\sigma$ confidence level. We choose a 2 keV particle ($4\sigma$ C.L.) in this paper to exaggerate the cosmological effects but which should not drastically affect structure formation on the scales considered. In our model dark energy is still described by a cosmological constant. In practice, $\Lambda$WDM is implemented by multiplying the initial power spectrum with a transfer function which truncates at smaller scales for higher energy particles. For a 2 keV particle the truncation scale corresponds to $0.15 h^{-1}$Mpc \citep{2001ApJ...556...93B}, well below the scales considered in this work.
\subsubsection{Quintessence}
\label{sec:quint}

Extended quintessence models describe dark energy through the evolution of a scalar field $\phi$ \citep{2015PhR...568....1J} whose Lagrangian is generically written as:

\begin{equation}
\label{Lagrangian}
L = \int d^4x \, \sqrt{ - g} \big( \, - \frac{1}{2} \partial_{\mu} \partial^{\mu} \phi +V(\phi)+m(\phi)\psi_m \bar{\psi}_m \big), 
\end{equation}
with a kinetic term, a potential term and an interaction term with the dark matter field $\psi_m$. The potential term can be chosen to produce the late time accelerated expansion we observe. In this work we use the Ratra-Peebles \citeyearpar{1988PhRvD..37.3406R} self-interaction potential:

\begin{equation}
V(\phi) = V_0 \phi^{-\alpha},
\end{equation}
where $\phi$ is in units of the Planck mass and $V_0$ and $\alpha$ are two constants that are fitted to observational data. 

The interaction term is designed to model the coupling mechanism between dark energy and dark matter. 
In this paper we study an `ordinary quintessence' model where $m(\phi) = m_0$ so that there is no direct coupling between the fields but the dark energy vacuum density has become time-dependent \citep{2004ApJ...605...21C}.

\subsubsection{Tweaked $\Lambda$CDM}

We also tested the dependence of the topological analysis on some cosmological parameters, namely the current dark energy density, $\Omega_{\Lambda}$, and the matter power spectrum normalization parameter, $\sigma_8$. These parameters are individually increased or decreased to the values given in Table~\ref{tab:params}. Increasing or decreasing the dark energy density moves the redshift of matter-dark energy equality from $z_{\Omega_{\Lambda}=\Omega_m}= 0.29$ to $ 0.60$ and $0.14$ respectively. As noted previously, comparing simulations with different cosmologies is difficult to do consistently. As the evolution of the scale factor depends on the underlying cosmology and parameters, the correspondence between two simulations at the same redshift requires careful interpretation. In this paper we consider the qualitative evolution of the Hermite spectra with redshift with the aim of distinguishing the general effects of a cosmological constant from those of a dynamical form of dark energy.

\begin{table}
	\centering
	\caption{Parameter values used in our simulations. $\Omega_{\Lambda}$ and $\sigma_8$ are individually reduced or increased, relative to $z=0$, but match with Planck observations at $z_\mathrm{CMB}$.}
	\label{tab:params}
	\begin{tabular}{  l | ccr } 
		\hline
		 & - & $\mathrm{nominal}$ & + \\
		\hline 
		$\Omega_{\Lambda}$ & 0.60 & 0.684 & 0.8095 \\
		$\sigma_8$ & 0.7 & 0.83 & 0.9 \\
		 $H_0$ & --- & 0.67 & --- \\
		$V_0$ & --- & $10^{-7}$ & --- \\
		$\alpha$  & --- & 0.143 & --- \\
		\hline
	\end{tabular}
\end{table}

\subsection{N-Body Simulations}
\label{sec: simul} 

\begin{figure}
	\includegraphics[width=\columnwidth]{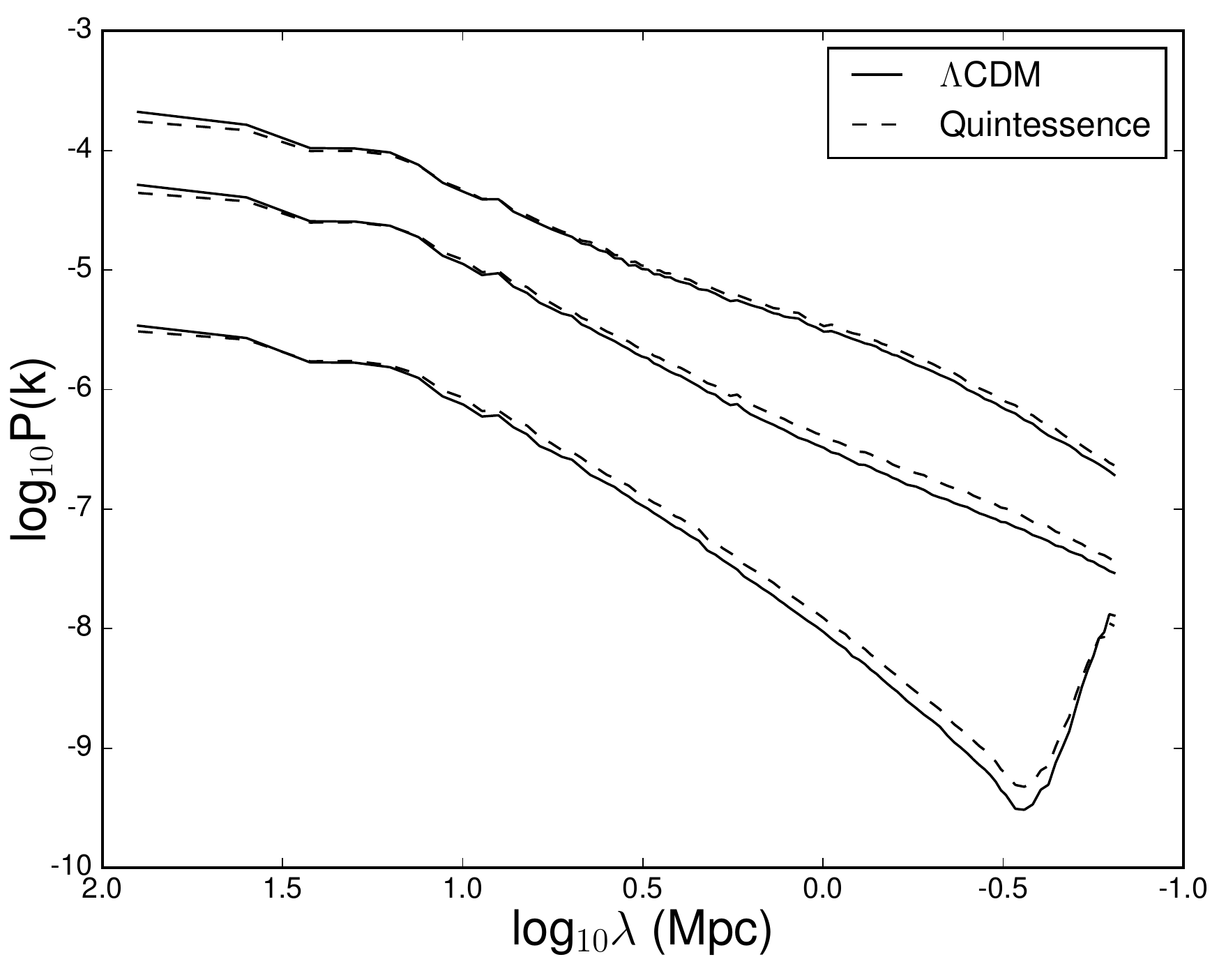}
    \caption{Power spectra for $\Lambda$CDM and a Quintessence model for $z= 0.00, 1.51, 9.00$ (from highest to lowest), showing tilting around a particular scale. The upturn for small scales at early times is due to shot noise.}
    \label{fig:powmesquint}
\end{figure}

Our hydronamical simulations are performed with P-GADGET-2, a modified version of the publicly available code GADGET-2 \citep{2005MNRAS.364.1105S} that is able to include non-standard dark sector physics. For details on the implementation, the reader is referred to an earlier paper \citep{2014MNRAS.439.2943C} which adapts the numerical implementation recipes in \citet{2010MNRAS.403.1684B}. The simulations we compare are produced with the same initial phase information so that individual objects and structures should appear at roughly the same position albeit with different density profiles depending on the power spectrum.  

In the case of non-standard cosmologies we must take care in generating the initial conditions, as there is some choice in what redshift we attempt to match cosmological parameter sets to observation. We could choose to match the parameters at $z=0$ or at $z_{CMB}$ as long as we only base our conclusions on what is being varied between simulation runs. In our model, we use cosmological parameters based on CMB observations by the Planck satellite interpreted in the context of a $\Lambda$CDM model, meaning that the redshift of the CMB is offset in our quintessence model, resulting in a modified early expansion history but giving the same $\sigma_8$ and expansion rate $H(a=1)$ today. The growth rates are calculated using $\mathrm{CMBEASY}$ and are used to produce GADGET format initial conditions using the publicly available N-GENIC1 MPI code. By this method, we hope to identify physical properties that behave differently amongst the models as they evolve. 

\begin{center}
\begin{tabular}{ l | c  r }
  \hline			
  Snapshot \# & Redshift(z) & Scale factor \\
\hline
  0 & 11.59 & 0.08 \\
\rowcolor{Gray1}
  1 & 9.00 & 0.10 \\
  2 & 6.94 & 0.13 \\
\rowcolor{Gray3}
 3 & 5.31 & 0.16 \\
 4 & 4.01 & 0.20 \\
\rowcolor{Gray5}
  5 & 2.98 & 0.25 \\
  6 & 2.16 & 0.32 \\
\rowcolor{Gray7}
  \textcolor{White}{7} & \textcolor{White}{1.51} & \textcolor{White}{0.40} \\
  8 & 1.00 & 0.50 \\
\rowcolor{Gray9}
  \textcolor{White}{9} & \textcolor{White}{0.59} & \textcolor{White}{0.63} \\
  10 & 0.26 & 0.79 \\
\rowcolor{Gray11}
  \textcolor{White}{11} & \textcolor{White}{0.00} & \textcolor{White}{1.00} 
	\label{tab:redshift}
\end{tabular}
\captionof{table}{Redshift and scale factor of the GADGET snapshots. Shaded rows are those used in all Hermite spectra figures throughout this paper.}
\end{center}

To check whether our results are robust to cosmic variance we also produced several realisations of the $\Lambda$CDM and Quintessence models. We calculated the power spectra using the POWMES code\footnote{powmes.0.2: http://www.projet-horizon.fr/article345.html} (results shown in Fig.~\ref{fig:powmesquint}). Due to the difference in early expansion history, Quintessence models will allow either more or less structure to form early (depending on the behaviour of the equation of state parameter $w=-p/\rho$) and hence tilt the normalised power spectrum around a characteristic scale.

\begin{figure}
\centering
	\includegraphics[width=\columnwidth]{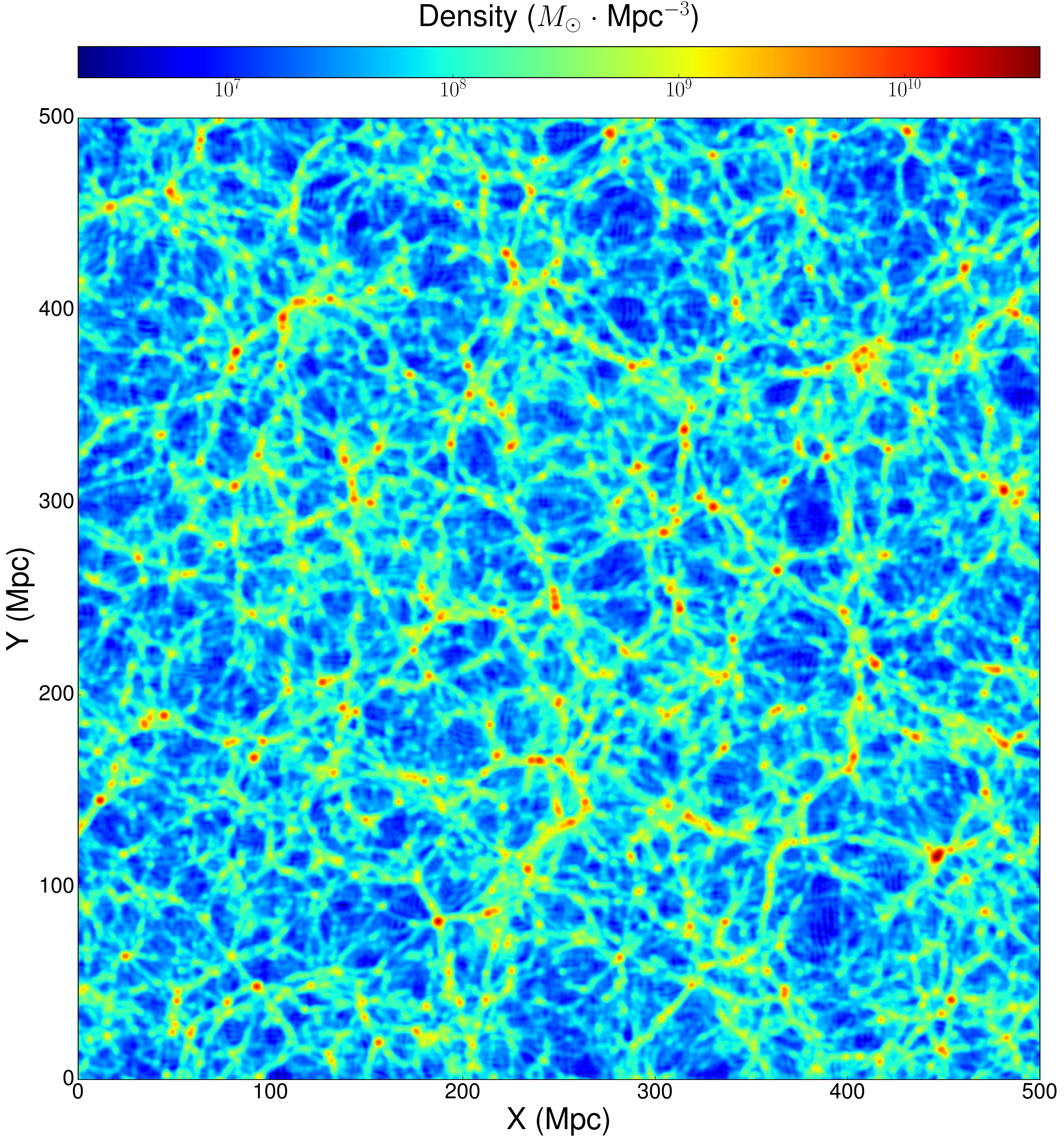}
    \caption{Density field at $z=0$ for a $\Lambda$CDM simulation, revealing the highly interconnected nature of Large-Scale Structure. Overdense (cluster-like) and underdense (void-like) regions are connected by filamentary structures in a vast `cosmic web'.}
    \label{fig:web}
\end{figure}

The accuracy of the genus statistic improves steadily with the volume of the data sample, so a large box size is crucial. We developed a highly-parallelized version of the genus calculation algorithm in order to increase the resolution of our density field and used the publicly available Fastest Fourier Transform in the West\footnote{fftw-3.3.4: fftw.org} to smooth the large datasets in parallel. We use a comoving box side length of $500 h^{-1}$Mpc with $1024^3$ particles and the density field resolution is $1024^3$ (for example, see Fig.~\ref{fig:web}). Each simulation is sampled over the entire cosmic evolution (see Table \ref{tab:redshift}) and we use Gaussian filters with a range of smoothing lengths from $1 h^{-1}$Mpc to $20 h^{-1}$Mpc. Calculating the genus of each sample uses 350 computing hours on the Raijin cluster of the National Computing Infrastructure\footnote{nci.org.au} in Australia.

\section{Results}
\label{sec:results}

\subsection{Meta-statistics}
\label{subsec:metastats}

\begin{figure}
	\includegraphics[width=\columnwidth]{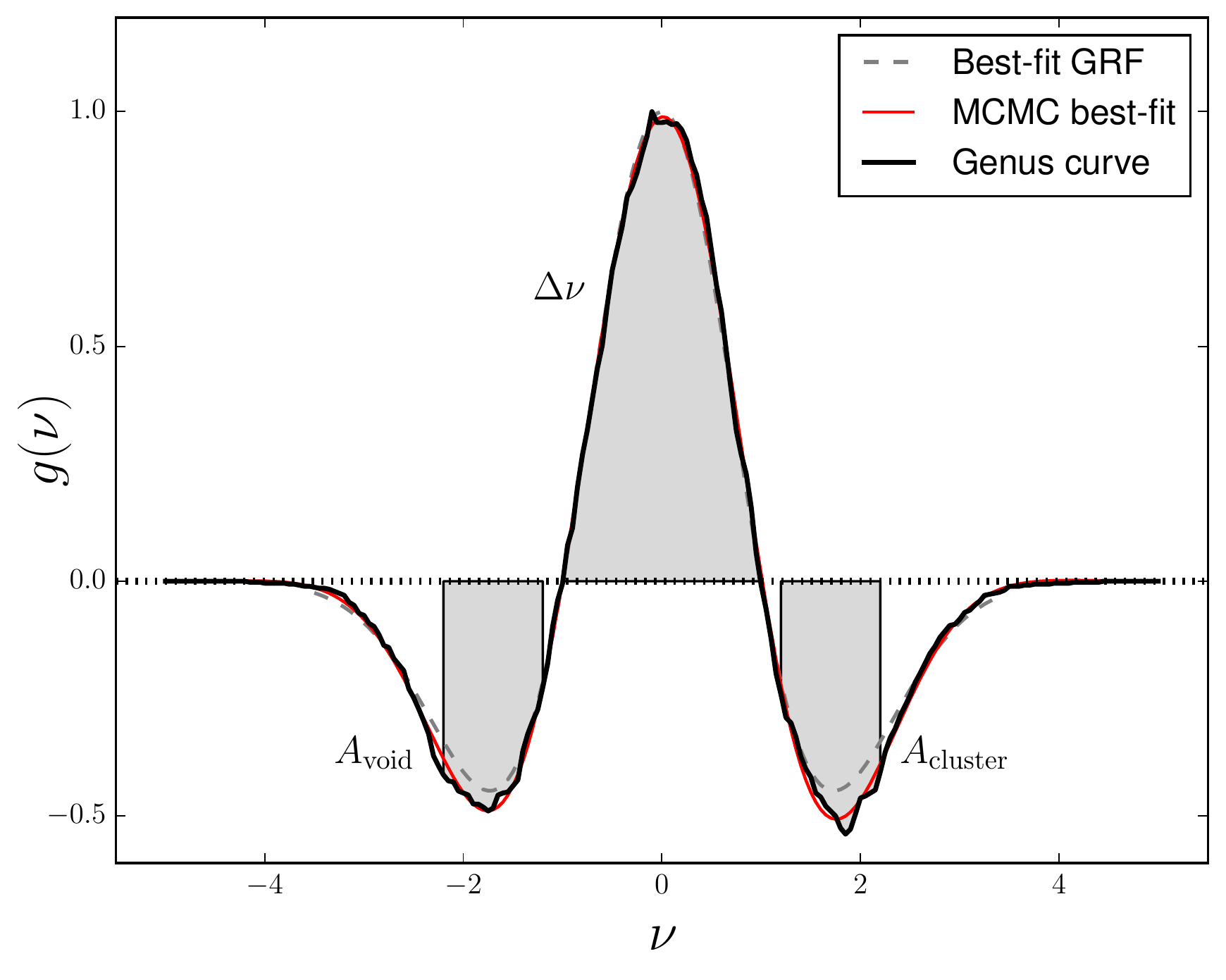}
    \caption{Normalized genus curve for a $z=0$ simulation smoothed at $10 h^{-1}$Mpc, our MCMC reconstruction (red) and the best-fit Gaussian Random Field (GRF). Integrating over the regions shown provides information about the abundance of void-like and cluster-like regions relative to a GRF.}
    \label{fig:genuscurves}
\end{figure}

Our suite of simulations was first analyzed in terms of the genus curves (see Fig \ref{fig:genuscurves}). At the small end of the wavelength range the Gaussian kernel is not particularly well resolved, resulting in more clumps and hence asymmetry in the genus curve toward overdensities. At the large end of the wavelength range, the box size limits the total number of holes and surfaces to less than 50 for any given density threshold and since the genus can only take integer values the curves become progressively less smooth. The Hermite spectra of these curves can show artifical noise at higher-order modes due to this effect, but the lower order modes should still be reliable. Due to these effects we believe the results are most reliable over the range $5 h^{-1}$Mpc to $15 h^{-1}$Mpc.

\begin{figure}
	\includegraphics[width=\columnwidth]{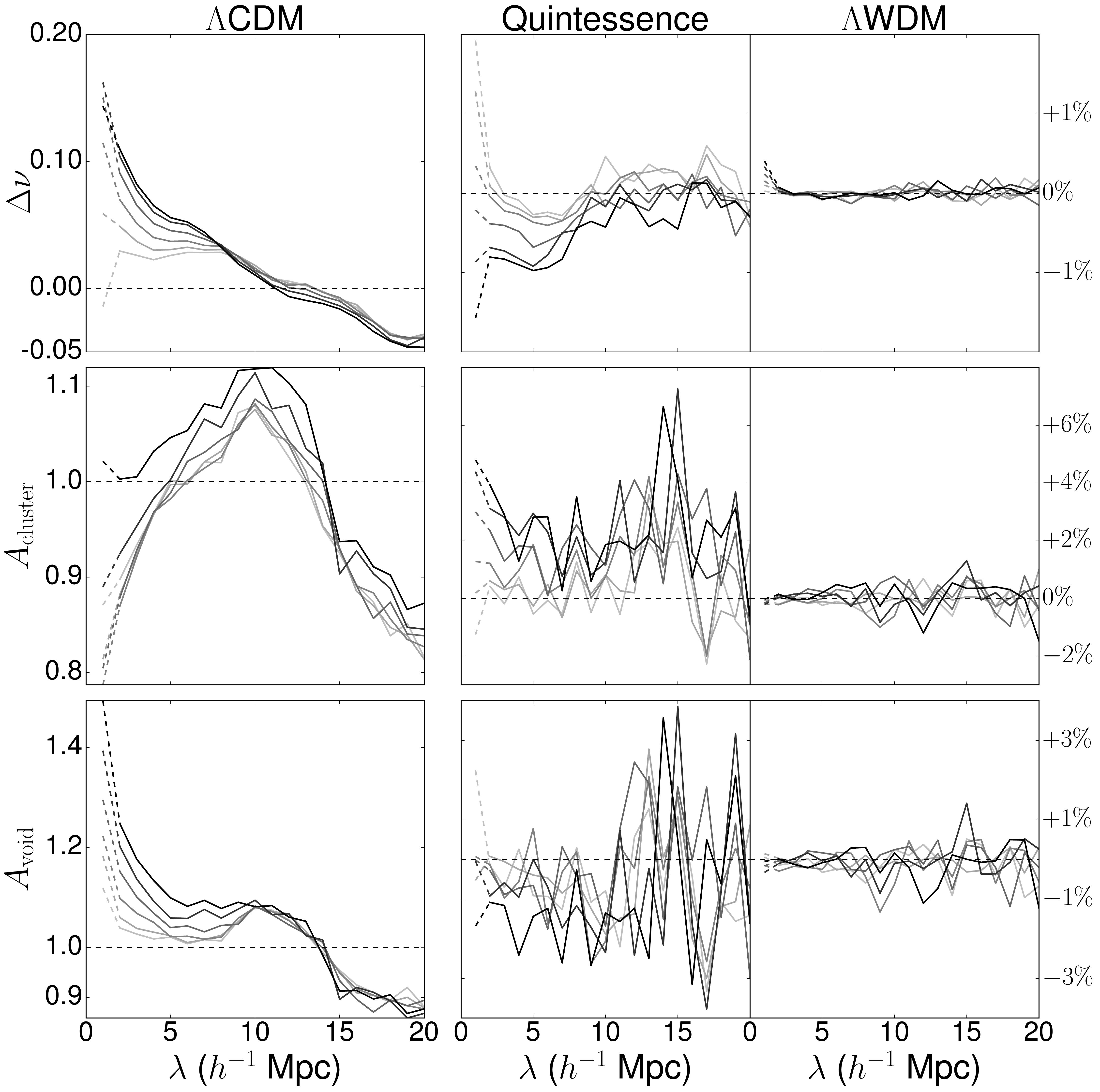}
    \caption{Meta-statistics ($\Delta \nu$, $A_{\mathrm{cluster}}$ and $A_{\mathrm{void}}$) for our cosmological models. Residuals after subtracting the non-standard models from a fiducial $\Lambda$CDM universe are also shown. The $\Lambda$WDM results are consistent with no difference; however, the Quintessence model produces $\sim2\%$ more clusters and $\sim1\%$ less voids.}
    \label{fig:stats}
\end{figure}

Given that the genus curve for a pure Gaussian Random Field (GRF) has an analytic formula, our computed genus curves can provide information about the non-Gaussian gravitational evolution through a simple comparison. It has become common to compute the following statistics proposed by \citet{1992ApJ...392L..51P, 2001ApJ...553...33P, 2005ApJ...633....1P}:

\begin{align}
\Delta \nu (\lambda) &= \frac{\int_{-1}^{1} \nu \, g(\nu;\lambda) \, d\nu}{\int_{-1}^{1} g_{\mathrm{GRF}}(\nu;\lambda) \, d\nu} \, , \\
A_v (\lambda) &= \frac{\int_{-2.2}^{-1.2} g(\nu;\lambda) \, d\nu}{\int_{-2.2}^{-1.2} g_{\mathrm{GRF}}(\nu;\lambda) \, d\nu} \, , \\
A_c (\lambda) &= \frac{\int_{1.2}^{2.2} g(\nu;\lambda) \, d\nu}{\int_{1.2}^{2.2} g_{\mathrm{GRF}}(\nu;\lambda) \, d\nu} \, .
\end{align}
$\Delta \nu$ is known as the \textit{genus shift} parameter and is positive or negative depending on how closely the topology resembles a clumpy or cellular structure respectively. $A_c$ and $A_v$ represent the abundance of overdense regions (clusters) and underdense regions (voids) respectively, relative to a GRF without gravitational evolution. The results of our meta-statistical analysis are shown in Figure~\ref{fig:stats}.

\begin{figure}
	\includegraphics[width=\columnwidth]{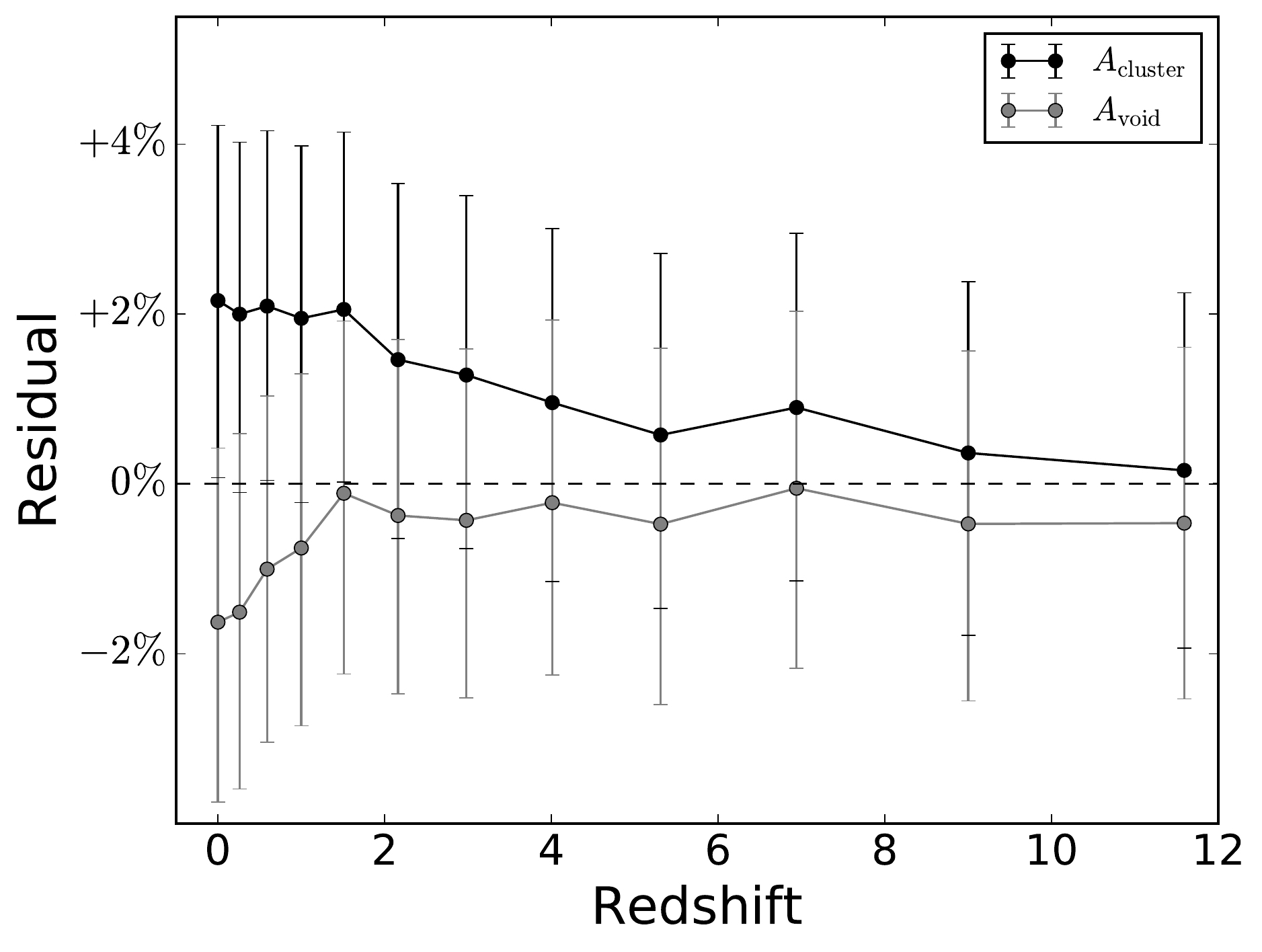}
    \caption{Residuals between Quintessence and $\Lambda$CDM as a function of redshift and assuming no wavelength dependence. Error bars are $1\sigma$ bounds from an MCMC analysis.}
    \label{fig:discrep_stats}
\end{figure} 
The trends for each model are similar; $\Delta \nu$ is positive below $\sim 10 h^{-1}$Mpc and negative above this scale, $A_c$ and $A_v$ are greater than 1 over the range $\sim5 - 15 h^{-1}$Mpc (higher cluster and void abundance than a GRF) and less than 1 otherwise. The discrepancies between the meta-statistics reveal that there is little difference between the $\Lambda$CDM and $\Lambda$WDM simulations at these length-scales, as is to be expected given that $\Lambda$WDM mostly affects the properties of individual halos and their substructures rather than the cosmic web. The quintessence model, however, has less positive $\Delta \nu$ below $10 h^{-1}$Mpc particularly for high redshifts, demonstrating a more cellular structure at small scales early in the history of the universe. The relative abundance of cluster-like regions is $\sim 2\%$ higher for the quintessence model while for void-like regions the abundance is $\sim 1\%$ lower (Fig.~\ref{fig:discrep_stats}). Both of these effects increase over cosmic time.

\subsection{Hermite spectra}
\label{subsec:hermspec}

\begin{figure}
	\includegraphics[width=\columnwidth]{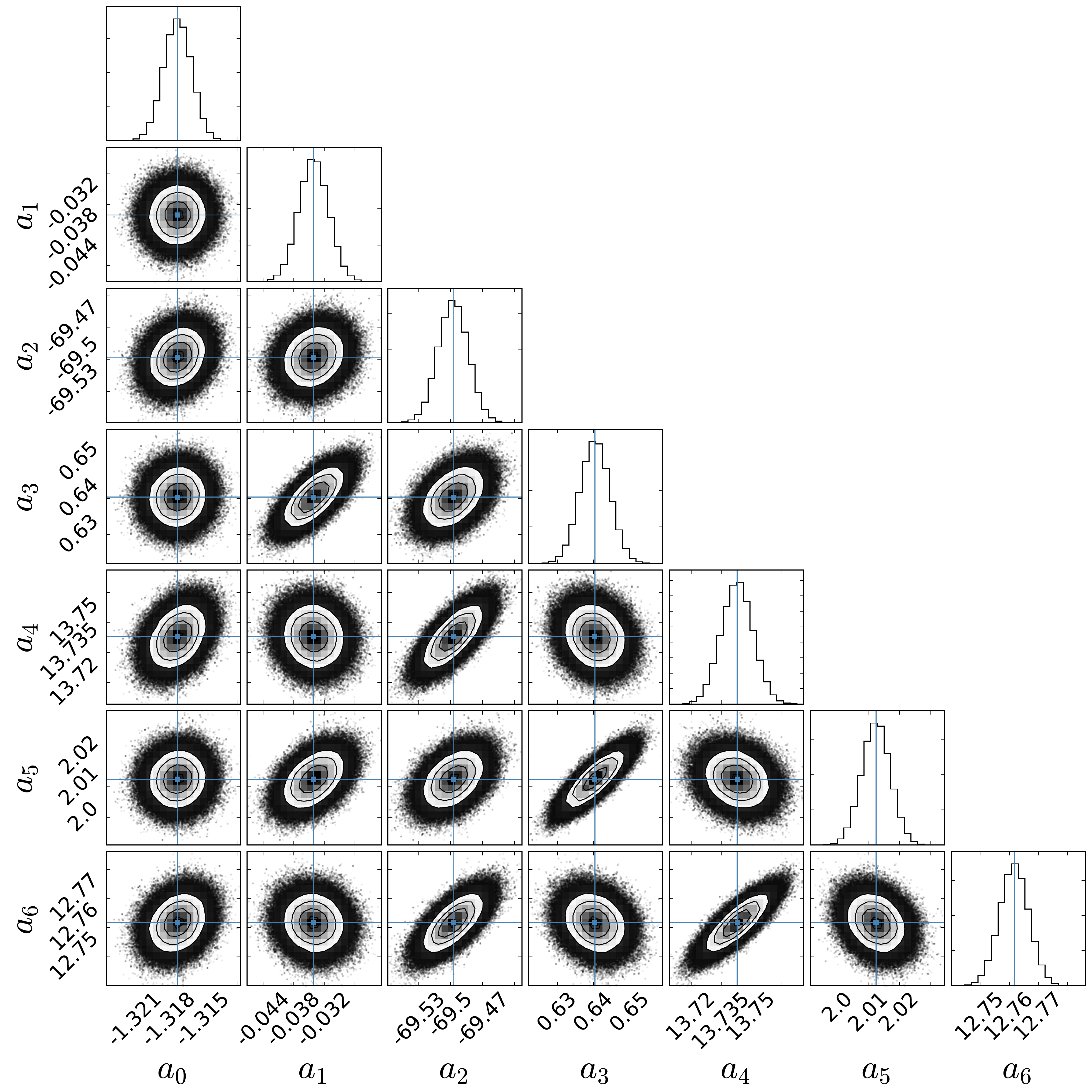}
    \caption{MCMC parameter estimation for the genus curve in Fig.~\ref{fig:genuscurves} as a percentage of the total contribution. The analysis reveals strong correlations between all the even modes and between all the odd modes, particularly for higher order modes. There is little to no cross-correlation between even and odd modes.}
    \label{fig:triangle}
\end{figure}

An example of the MCMC results is shown in Fig.~\ref{fig:triangle}. These reveal that there are strong correlations in the coefficients between all the even modes and between all the odd modes. The correlations get stronger for higher order modes ($m=4,5,6$), reflecting the difficulty of disentangling the higher order modes from the more fundamental ones, particularly the dominant $a_2$. The even modes are not correlated with the odd modes at all, so the signal of asymmetries in the curve should be strong.

\begin{figure*}
	\includegraphics[width=0.8\textwidth]{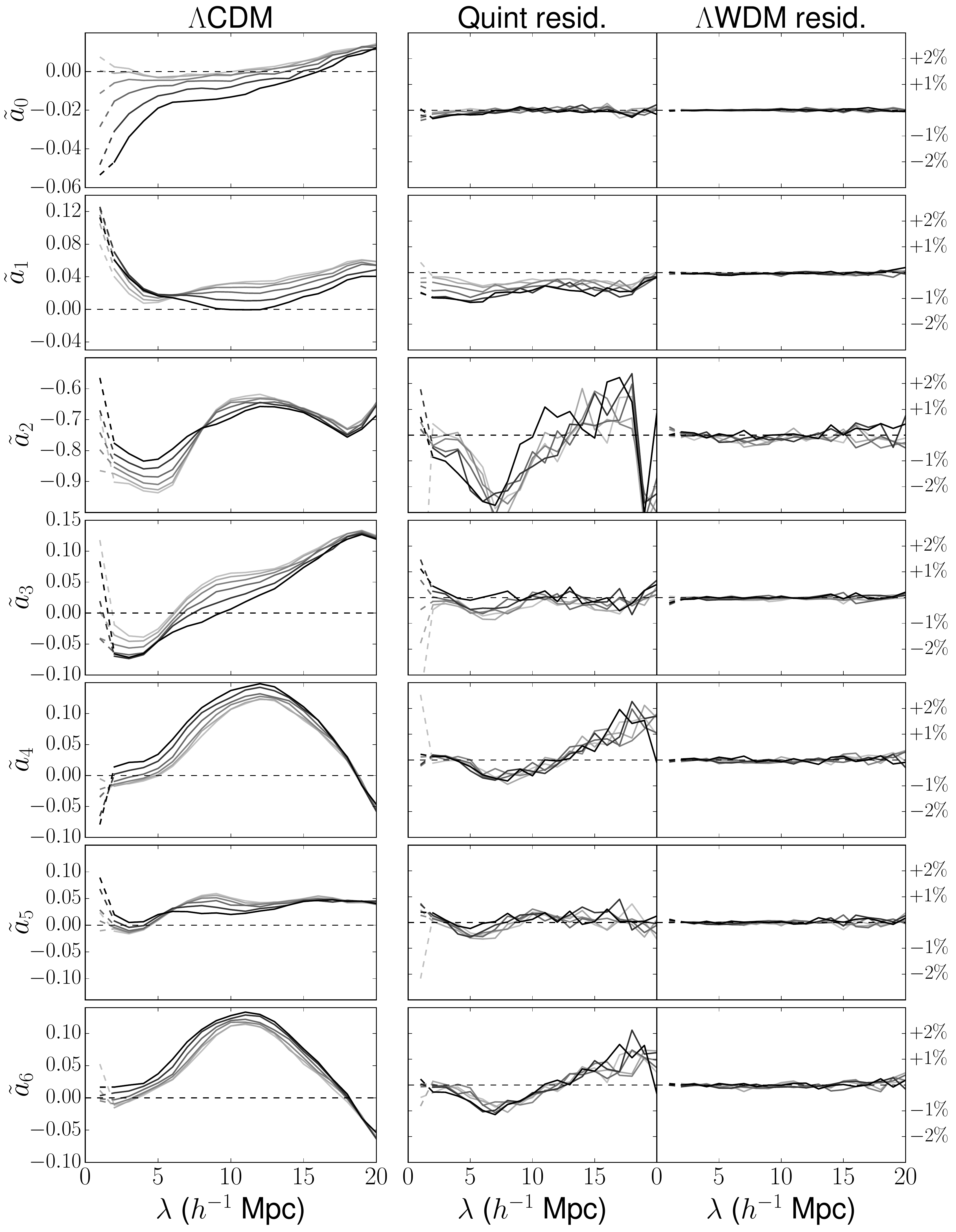}
    \caption{Hermite spectra for modes $m$=$0,\ldots,6$. Darker curves represent smaller redshifts. The spectra evolve smoothly in time and are roughly consistent between the models, with the exception of $\tilde{a}_1$ which is lower for the Quintessence model, indicating more clusters and less voids.}
    \label{fig:herm}
\end{figure*}

Our results for the three models are shown in Figures~\ref{fig:herm}. Darker curves represent smaller redshifts with the darkest corresponding to $z=0$.  We also show the residuals by subtracting the Hermite spectra from the fiducial $\Lambda$CDM model. The dominant $a_2$ mode contributes about $80\%$ to the genus curves, with each of the other modes contributing $<10\%$. All the curves evolve smoothly with wavelength and redshift.

For all the modes there is essentially no difference between the $\Lambda$CDM and $\Lambda$WDM models, in agreement with the metastatistics analysis and theoretical expectations. The Quintessence model produces interesting differences in almost all the modes. While the $a_0$ mode evolution remains the same, the $a_1$ contribution slowly decreases relative to the fiducial model in a nearly scale-independent manner. This negative shift in the linear mode will produce an asymmetry in the genus curves and weight them toward more clusters and fewer voids, in qualitative agreement with our metastatistics analysis, though the smooth evolution and acceleration of this process is much clearer in the Hermite spectra. The differences in the $a_2$ mode are less clear cut - there is some evidence of redshift evolution but it appears to be more noisy and scale-dependent.

The differences in the higher order modes ($m$=$3,\ldots,6$) do not show substantial redshift evolution but are wavelength dependent. There is little difference between the models below $10 h^{-1}$Mpc. Above this scale, the odd modes show a negative deviation and the even modes show a positive deviation of very similar magnitude. 

There is also an interesting feature in the data points at $1 h^{-1}$Mpc, in that they appear to `splay'. Although these measurements are approaching the cell size of our density field, the strong redshift dependence suggests that there are cosmological effects at the lower wavelength limit worth further investigation with better resolution.

\begin{figure*}
	\includegraphics[width=0.7\textwidth]{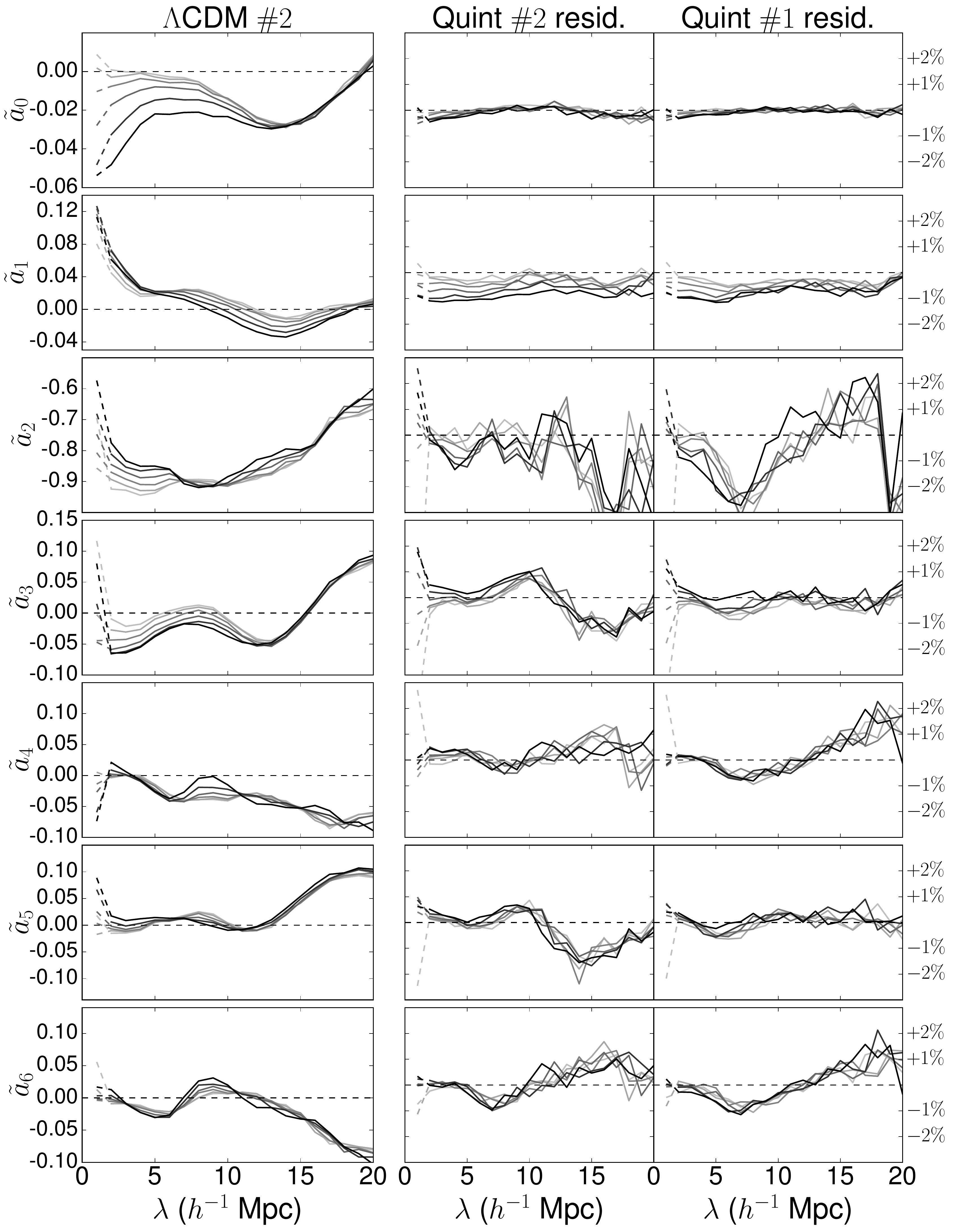}
    \caption{Hermite spectra for modes $m$=$0,\ldots,6$ for a second seeding of the initial conditions. Trends are similar to those in Fig~\ref{fig:herm}, particularly the wavelength-independence of the $a_1$ residuals and weak wavelength dependence of higher order modes.}
    \label{fig:cosmicvar}
\end{figure*}

We test the sensitivity of our topological analysis to cosmic variance by producing parallel simulations of each cosmology with different initialisation seeds. The power spectra of the alternate simulations remain unchanged except at low $k$. The comparison of each simulation is shown in Fig~\ref{fig:cosmicvar}, with the $\Lambda$WDM model omitted since the results are again nearly identical to the fiducial model. We find that the scale and redshift dependence of the Hermite coefficients is altered, though remains of the same order and qualitatively similar shape. This means we cannot interpret the individual curves as an observable result if the simulations are not first constrained by galaxy catalogues and other maps of large-scale structure. However, the differences that the Quintessence model produces are consistent between the two realisations. In particular, the scale-indepedent reduction in $a_1$ is reproduced. Similar trends in the higher order modes are also observed, though the difference in $a_4$ is slightly muted in the second realisation, possibly because $a_4$ contributes less overall. Cosmic variance is a significant factor in our results, but simultaneously the robustness of our results is strengthened by examining independent simulations. 

\begin{figure*}
	\includegraphics[width=\textwidth]{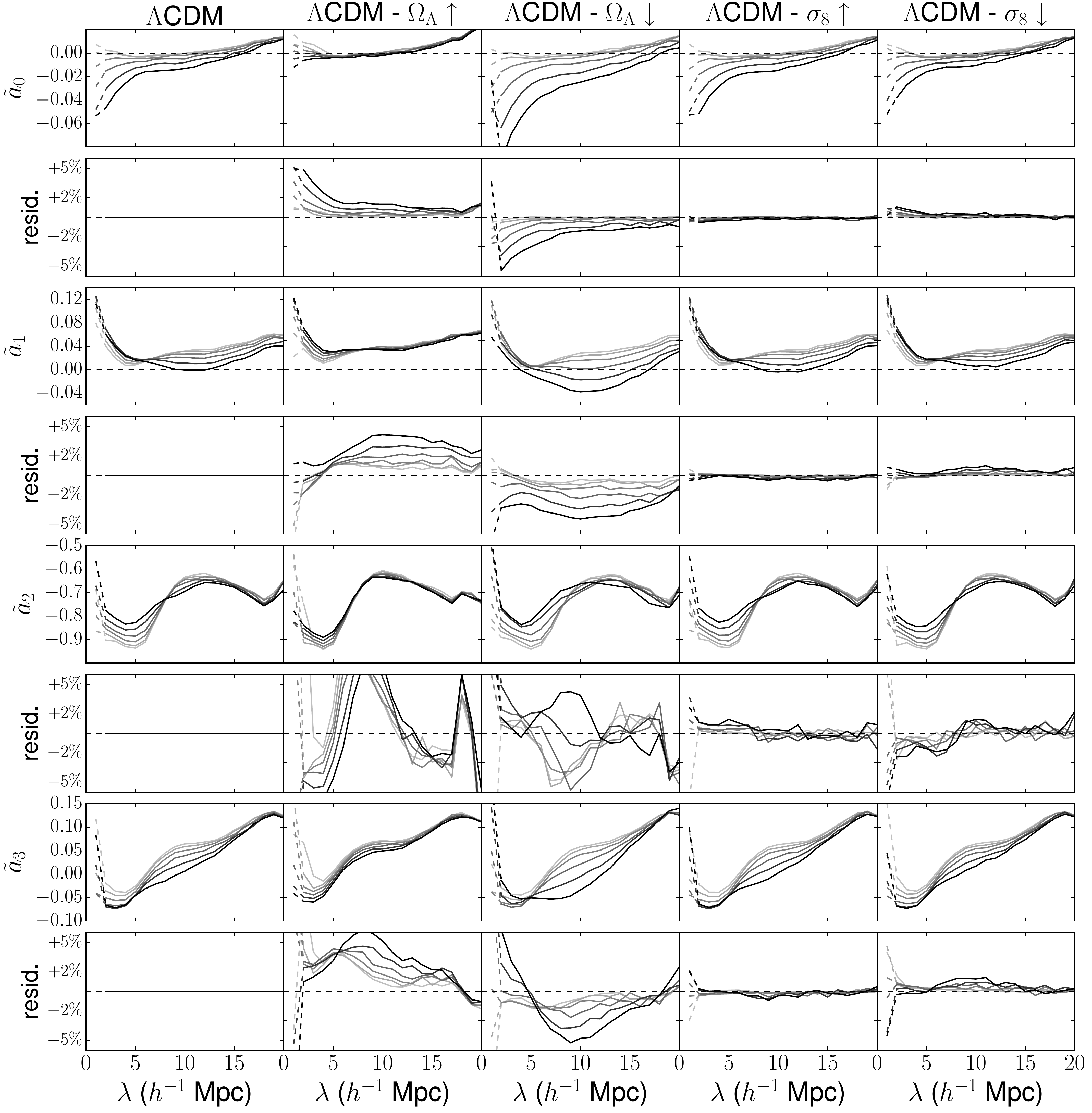}
    \caption{Hermite spectra for modes $m$=$0,1,2,3$ for several realisations with different parameter values as given in Table~\ref{tab:params}. Note that the magnitude scale on the residuals plots has increased from $2\%$ to $5\%$.}
    \label{fig:deltaparam}
\end{figure*}
The effects of modifying the cosmological parameters of a $\Lambda$CDM model are shown in Fig~\ref{fig:deltaparam}. We focus on $\Omega_{\Lambda}$ and $\sigma_8$, which alter the expansion history and the scale factors at which scales go nonlinear. We see that topology is strongly effected by $\Omega_{\Lambda}$ and less so by the normalisation of the matter power spectrum given by $\sigma_8$. 
A high value for the dark energy density results in almost no redshift evolution of any Hermite spectra beyond $10 h^{-1}$Mpc, suggesting that structure on the largest scales is frozen-in at high redshift, with modest evolution on smaller scales. Conversely, a lower dark energy density accelerates the redshift evolution and produces differences in the genus curve which are approximately equal but opposite. The two cosmological parameters also negatively correlate, with increases in $\sigma_8$ producing shifts in the curves in the same direction as decreases in $\Omega_{\Lambda}$. In contrast to the Quintessence model, modes other than $a_1$ also differ to the fiducial model by similar magnitudes and across all wavelength scales. The splaying of the curves at $1 h^{-1}$Mpc is also observed in these simulations.

\section{Discussion}
\label{sec:discussion}
Understanding the differences identified in the Hermite coefficients for the cosmological realisation presented in this paper are not necessarily straight forward, but we can interpret these features in terms of their impact on large scale structure. There are no cosmological differences between the $\Lambda$CDM and $\Lambda$WDM models, at least at the energy scale of 2 keV and down to scales of $1 \, h^{-1}$Mpc.  
At the scales we sample, it is clear that topology is insensitive to the mass-scale of the DM particle.

The Quintessence model produces topological differences that can be detected in multiple ways. Our metastatistical analysis indicates that Quintessence should have a higher abundance of cluster-like regions and lower abundance of void-like regions relative to a pure Gaussian Random Field, but the results are noisy. Analysis of the Hermite spectra reveals a similar effect in that the odd modes are typically reduced relative to the $\Lambda$CDM case with a clear dependence on redshift and more weakly on smoothing length. These results are fairly robust to the effects of cosmic variance, but the exact shape of the Hermite spectra depend on the volume being studied.

The differences produced in the Quintessence model are characteristically different to those produced by tweaking the cosmological parameters of a $\Lambda$CDM model. While simulations produced with a slightly reduced value for the current dark energy density, $\Omega_{\Lambda}$, will lead to a similar reduction in the linear Hermite mode $a_1$, they also reduce the constant Hermite mode $a_0$ which is not observed in the Quintessence case. They also do not reproduce the same wavelength dependence in the shifts of the higher order modes. There is some ambiguity in comparing simulations with different cosmologies and expansion histories, but we have at least identified a number of qualitative differences.

Our results indicate that in principle there are measurable differences in the topology of large-scale structure for even minor changes in the underlying dark sector model. 
To compare to observations, we would need to use galaxy surveys with a large survey volume and significant depth to measure the Hermite spectra accurately and map their redshift dependence.
The effects of spatially-incomplete samples, flux-limited observation techniques and experimental noise will undoubtedly make signatures of dark sector physics in large-scale structure harder to identify than in simulations. 


\section{Conclusions}
\label{sec:conclusion}

We have studied the topology of large-scale structure in several cosmological models using the genus statistic and its corresponding Hermite spectra. We have found that a fiducial $\Lambda$CDM model and a $\Lambda$ Warm Dark Matter model with particle energy 2 keV produce nearly identical results at super-galactic scales, in line with theoretical predictions \citep{2011PhRvD..84f3507S}. We also studied a Quintessence model that includes a dynamical form of dark energy rather than a cosmological constant. Since our simulations have no baryons, the Quintessence model differs from the $\Lambda$CDM model only in its expansion history. We found several differences including higher cluster abundances and lower void abundances, a redshift-dependent reduction in the linear Hermite mode contribution and soeme weakly wavelength-dependent effects in higher order modes. These differences are moderately robust to cosmic variance and are characteristically different from the effects of tweaking several parameters of a $\Lambda$CDM model.

Given the similarity of these dark-matter only models, we expect to see more pronounced differences once gas physics is included in the simulations. Our next step is to study the models including baryons over a wide parameter range that includes the strength of the dark sector coupling and the energy of the WDM particle. The question of the detectability of the topological signatures of non-standard cosmologies in current and future large sky surveys has yet to be addressed; we leave this for a future contribution. 

\section*{Acknowledgements}

AW is supported by an Australian Postgraduate Award. PJE acknowledges funding from SSimPL programme and the Sydney Institute for Astronomy (SIfA), DP130100117 and DP140100198, ARC Discovery Projects awarded to GFL and CP. This research was undertaken on the NCI National Facility in Canberra, Australia, which is supported by the Australian Commonwealth Government, with resources provided by Intersect Australia Ltd and the Partnership Allocation Scheme of the Pawsey Supercomputing Centre. The authors acknowledge the University of Sydney HPC service at The University of Sydney for providing HPC resources in the form of the Artemis Supercomputer\footnote{sydney.edu.au/research\_support/hpc} that have contributed to the research results reported within this paper. 











\bsp	
\label{lastpage}
\end{document}